\documentclass[
    prl,twocolumn,
    superscriptaddress,
    amsmath,amssymb,aps,
    floatfix,longbibliography,10pt
]{revtex4-2}

\usepackage{silence}
\usepackage{etoolbox}
\usepackage{xr-hyper}
\usepackage{hyperref}

\hypersetup{colorlinks=true,breaklinks=true,linkcolor=red,citecolor=blue,urlcolor=blue,  linktocpage=true, pdfborder={0 0 0}}

\makeatletter
\newcommand*{\addFileLabels}[1]{%
  \temp@aux{#1}%
}
\newcommand*{\temp@aux}[1]{%
  \@input{#1.aux}%
}
\makeatother

\usepackage[utf8]{inputenc}
\usepackage{amsmath}
\usepackage{amssymb}
\usepackage{graphicx}
\usepackage{babel}
\usepackage{mathrsfs}
\usepackage{amsfonts}
\usepackage{epstopdf}
\usepackage{slashed}

\usepackage{multirow}

\makeatletter
\patchcmd{\@bibdataout@aps}{author="08"}{author="48"}{}{}
\makeatother

\begin{document}
\title{Topologically shadowed quantum criticality: A non-compact conformal manifold}

\author{Tianyao Fang}
\affiliation{Department of Physics, The Chinese University of Hong Kong, Sha Tin, New Territories, Hong Kong, China}

\author{Weicheng Ye}
\affiliation{Department of Physics and Astronomy, University of British Columbia, 6224 Agricultural Road, Vancouver, BC, V6T 1Z1, Canada}

\author{Zheng-Cheng Gu}
\affiliation{Department of Physics, The Chinese University of Hong Kong, Sha Tin, New Territories, Hong Kong, China}
\affiliation{The State Key Laboratory of Quantum Information Technologies and Materials, The Chinese University of Hong Kong, Shatin, New Territories, Hong Kong, China}

\author{Fei Zhou}
\affiliation{Department of Physics and Astronomy, University of British Columbia, 6224 Agricultural Road, Vancouver, BC, V6T 1Z1, Canada}

\begin{abstract}

We propose a family of topological quantum critical points (tQCPs) separating non-invertible chiral topological orders in $(2+1)$ dimensions. We conjecture that these tQCPs can be captured by a family of scale-invariant field theories forming a non-compact scale-invariant manifold. A central feature of our proposal is topological shadowing: the underlying critical theory is rigorously constrained by the global topological data of the two adjacent gapped phases. These theories can be further projected into quantum field theories with universal non-local structures. The non-local coupling, however, controls dynamical observables along the scale-invariant manifold, including the VEV of Wilson loops and their exchange phase around the two non-contractible cycles of a torus.
%Specifically, we show that the $U(1)$ symmetric critical point characterized by a topological angle $\Theta_{\text{cft}}$, defined by a commutator between two Wilson loop operators on a torus, is uniquely determined by the braiding angles $\Theta_{1,2}$ of the adjacent gapped phases via the relation $\Theta_{\text{cft}}^{-1} =\frac{1}{2}[\Theta_1^{-1} + \Theta_2^{-1}]$. 
Despite the non-locality, our renormalization group calculations (up to two-loop order) strongly suggest that the theory shall maintain exact scale invariance. This supports, without supersymmetry, a continuous family of scale-invariant fixed points, which are further substantiated from a holographic perspective. %manifold of fixed points that naturally becomes a conformal manifold.
%when a local structure is further enforced.
\end{abstract}

%\date{\today}
\maketitle

\textit{Introduction}---The classification of quantum phases has evolved from the Landau-Ginzburg paradigm to topological orders~\cite{Wen_book, Kitaev_Anyons}. Recently, this frontier has expanded further with the discovery of generalized symmetries, particularly non-invertible symmetries (or categorical symmetries), which have provided deep insights into topological boundaries \cite{Gaiotto_2015, McGreevy_Review, Shao_NonInvertible, Cordova_2022}. While classifying gapped phases is increasingly sophisticated, understanding the continuous topological quantum critical points (tQCPs) connecting them remains a formidable challenge. Unlike standard Landau transitions driven by local order parameters, tQCPs typically involve fractionalized degrees of freedom and emergent gauge fields \cite{Senthil_DQCP, Grover_tQCP,Sachdev_book,Wen_2017}. An essential question is: given two distinct topologically ordered phases, {\em can we systematically determine the nature of the critical point separating them?} Furthermore, does the dynamics at such critical points allow for continuous deformations, or do isolated fixed points represent them just like in conventional quantum critical points (QCP)? Although recent development of generalized symmetry and holographic scenario impose strong constraints on tQCPs, there is still lacking concrete understanding, especially in higher dimensions.   

In this work, we address these questions by proposing a theoretical framework for topologically shadowed tQCPs in $(2+1)D$. We focus on transitions between non-invertible chiral topological orders characterized by distinct chiral central charges. 
%In the standard renormalization group (RG) paradigm, a quantum critical point represents an infrared (IR) unstable fixed point; a relevant perturbation, such as a fermion mass, drives a RG flow toward the gapped topological phases. 
The dynamical renormalization group (RG) flow naturally proceeds from a critical point to a gapped phase, the principle of 't Hooft anomaly matching~\cite{tHooft1980,WenAnomaly2013} dictates that the global topological anomalies must be strictly conserved along the entire flow.
The topological data of the gapped topological phases---such as chiral central charges and anyon topological spins---are robustly quantized and conceptually well-understood. We exploit this conservation 
%to reverse the standard direction of logical inference: rather than %deducing the properties of the gapped phases from a given critical %theory, we 
by utilizing the rigid topological data of both adjacent gapped phases to strictly constrain the parameters of the gapless critical point. We term this mechanism "Topological Shadowing", where the gapped topological phases effectively cast a structural shadow at the gapless critical point. Specifically, we consider an effective field theory of $N_f$ flavors of massless Dirac fermions coupled to a level $k$ Chern-Simons gauge field. We demonstrate that the defining parameters of this critical theory ($k$ and $N_f$) can be uniquely constrained by the chiral charges ($c_1, c_2$) of the adjacent gapped phases. 
%This establishes a rigorous correspondence between the bulk topology and the critical dynamics.

Most remarkably, our analysis of the quantum dynamics reveals a structure rarely seen in $(2+1) D$  non-supersymmetric theories. Upon integrating out the gapless fermions, the effective gauge theory acquires a universal non-local structure in its quadratic term. Despite this non-locality, we find that the theory maintains exact scale invariance through explicit RG calculations up to the two-loop order. The beta functions for both the gauge coupling and the emergent non-local term vanish identically. This implies that these tQCPs do not exist as isolated points but form a non-compact scale-invariant manifold in the parameter space. This finding presents a $(2+1)D$ analogue to the Luttinger liquids in $(1+1)D$ \cite{Haldane_Luttinger}, but realized here in a topological gauge theory context without invoking supersymmetry \cite{Seiberg_SUSY_Manifolds}. Along this manifold, $\gamma_\psi$ and, in the large-$k$ approximation, the Wilson-loop amplitude and ordered exchange phase vary continuously with $\lambda$, while the discrete topological matching constraints remain fixed.

%We further show that along this manifold while the fermion anomalous dimension $\gamma_{\psi}$ varies continuously, its universal topological signature (characterized by a topological angle $\Theta_{\text{cft}}$ in the Wilson loop braiding) remains invariant, and is fully determined by the braiding statistics of the adjacent gapped phases.

%The remainder of this paper is organized as follows. In Sec.~\ref{sec:eEFT}, we construct the extended effective field theory and derive the constraints on $k$ and $N_f$. In Sec.~\ref{sec:manifold}, we present the renormalization group analysis and the evidence for the scale-invariant manifold. We further discuss the Wilson loop algebra and the topological angle. Technical details regarding the two-loop beta function calculations and the Wilson loop algebra are provided in the Appendices.

\textit{Extended Effective Field Theory}---  We begin with a $(2+1)$-dimensional extended effective field theory (eEFT) describing a critical point between distinct topological phases. The theory consists of $N_f$ flavors of massless Dirac fermions $\psi_i$ coupled to a $U(1)$ gauge field $a_\mu$ with a Chern-Simons term at level $k$. To capture the full quantum dynamics at the critical point, 
%including potential renormalization effects, 
we generalize the standard action to include a gauge invariant non-local term which naturally emerges if the fermions are massless. 
%The Lagrangian in Euclidean signature is specified as:
\begin{align}
\mathcal{L}_{\text{eEFT}} &= \frac{i k}{4\pi} \epsilon_{\mu\nu\lambda} a_\mu \partial_\nu a_\lambda + \frac{\lambda}{2} a_\mu \hat{\Pi}_{\mu\nu} a_\nu \nonumber \\
&+ \sum_{i=1}^{N_f} \bar{\psi}_i \gamma_\mu (\partial_\mu - i g a_\mu) \psi_i,
\label{eq:lagrangian}
\end{align}
where g is the coupling constant in the unit of charge and $\gamma_\mu$ are the $(2+1)$D Dirac matrices. The non-local operator $\hat{\Pi}_{\mu\nu}$ in the quadratic term is defined in momentum space as the transverse projection operator:
\begin{equation}
\hat{\Pi}_{\mu\nu}(q) = |q| \left(\delta_{\mu\nu} - \frac{q_\mu q_\nu}{q^2}\right),
\label{eq:non_local_term}
\end{equation}
which respects $U(1)$ charge conservation $q_\mu \Pi_{\mu\nu}(q) = 0$.
In the UV limit of the effective theory, the bare parameter $\lambda$ may be zero. However,  in realistic materials with a hierarchy of Fermi velocities, integrating out a subset of ‘fast’ massless Dirac fermions—which are kinematically decoupled from the lowest-energy modes—precisely yields this non-local polarization background for the emergent gauge field~\cite{patrick2006, Pisarski1984, Hermele_2005}.  Since the effective coupling $\lambda$ is inversely proportional to the Fermi velocity $v_F$ of these integrated-out excitations \cite{DasSarma2011, Pizarro2019}, the required large velocity separation between the fast and slow sectors ensures that $\lambda$ is typically small. It is crucial to note that this emergent non-local term scales linearly with momentum $q$, exactly like the topological Chern-Simons term. They have the same conformal dimension, appearing equally relevant in the renormalization group analysis. This is in sharp contrast to the standard local Maxwell term ($\sim q^2$), which becomes irrelevant in the infrared limit compared to the $q$-term. 
%For this reason, the non-local structure is not merely a correction but a defining feature of the critical theory, and we include it explicitly in our eEFT.

\textit{Topological shadowing and anomaly matching}-- The parameters $\{k, N_f, g\}$ of this critical theory are not free parameters. They are rigidly constrained by the topological nature of the two adjacent gapped phases. We term this constraint mechanism topological shadowing. 
In the eEFT for the charge $U(1)$ sector, a transition is simply tuned by a fermion mass term $m \sum_i \bar{\psi}_i \psi_i$. The two gapped phases correspond to $m > 0$ and $m < 0$.
For the gapped phases with topological orders of $U(1)_{k^{\pm}_G}$, the levels $k_G^{\pm}$ are quantized integers, i.e. $k^{\pm}_G \in \mathbb{Z}$ where the superscripts $\pm$ further indicate the two gapped phases with $m > 0$ and $m < 0$, respectively.

Deep in the gapped phases, the fermions can be integrated out. A single massive Dirac fermion of mass $m$ contributes a level $\frac{1}{2} \text{sgn}(m)$ to the Chern-Simons term due to the parity anomaly \cite{Redlich_Parity, Niemi_Seminoff}. Consequently, the effective Chern-Simons level in the eEFT theory that shifts from the bare value $k$ are directly constrained by the topological data of the gapped phases $k^{\pm}_G$ via: 
\begin{eqnarray}
    k \pm \frac{N_f}{2}=k_{\text{G}}^{\pm}.
    \label{eq:shadowing0}
\end{eqnarray}

We anticipate that the topological response calculated from the eEFT describing the phase transition must match the inherent topological data of the two gapped phases, similar to the principle behind the duality web\cite{PhysRevLett.47.1556,Seiberg2016Web,Hsin:2016blu}.
This is also reminiscent of the 't Hooft anomaly matching condition\cite{tHooft1980} if we treat these topological data as the data for the \emph{generalized symmetries} of the phase diagram\cite{Cordova_2022}. 
%The principle of 't Hooft anomaly matching\cite{tHooft1980,WenAnomaly2013} dictates that the topological response calculated from an eEFT theory or IR theory must match the inherent topological data of the gapped phases. 

To be more illustrative, we further consider a specific
conformal embedding of the charge $U(1)$ Chern-Simons sector in a $G=U(1) \times SU(n)$ WZW theory with central charges
$c_{1,2}=\frac{1}{2} \nu_{1,2} \in \mathbb{Z}$ and $\nu=2n, n=0,1,2...$.
Physically, this can be practically achieved in chiral superconductors with central charges $c_{1,2}$ but further enriched with $U(1) \times SU(n)$ gauge symmetry.
%\footnote{For the standard D-class states (i.e.unenriched), $\nu_{1,2}=2c_{1,2}$ is also the Chern-number specifying the global band topologies.}.
%Now consider a phase transition between a topological phase with chiral central charge $c_1$ to another with $c_2$ (we assume $c_2 > c_1$ without loss of generality). 
To embed the $U(1)_{k_G}$ theory in such chiral phases, we need to specifically demand that $k_G^{\pm}$, {\em the levels of the charge $U(1)$ Chern-Simons sector} for two gapped bulk phases 
have to match, respectively, the numbers of chiral edge modes in the {\em unenriched limit}, i.e. the central charges $c_{1,2}$\footnote{This is effectively an UV-IR anomaly matching. Note that the central charge in the eEFT of the $U(1)_k$ charge-sector alone is $c_{U(1)}=1$. Although the rest of central charges, $[c_{1,2}-1]$, are distributed among the $SU(c_{1,2})_1$ sectors, these {\em charge-neutral sectors} don't contribute to the $U(1)$ chiral mode under consideration here.}.

This 't Hooft anomaly matching in two distinct gapped phases via a holographic principle leads to an identity which further relates $k, N_f$, the parameters in the eEFT in the IR limit to the UV topological data of $c_{1,2}$;

\begin{equation}
k - \frac{N_f}{2} = c_1, \quad
k + \frac{N_f}{2} =c_2. \label{eq:matching}
\end{equation}
Solving these coupled equations, we derive the fundamental constraints defining the topological shadow:
\begin{equation}
k = \frac{1}{2}(c_1 + c_2), \quad N_f = c_2 - c_1, \quad c_{1,2}\in \mathbb{Z}.
\label{eq:shadowing_result}
\end{equation}
Here, we have assumed for simplicity that the fermions carry the fundamental gauge charge ($g=1$). In a more general scenario where the gapless fermions carry a fractional charge $g = 1/p$ (with $p \in \mathbb{Z}$), the condition for the number of flavors generalizes to $N_f = p^2 (c_2 - c_1)$ to compensate for the reduced coupling strength.

An immediate consequence of Eq.~(\ref{eq:shadowing_result}) arises from the fact that we consider a compact $U(1)$ gauge group. This compactness implies that the chiral charges $c_{1,2}$ characterizing the topological phases have to be integers. Consequently, the Chern-Simons level $k$ is generally a half-integer (also see Eq.~(\ref{eq:shadowing0})). This result is in full agreement with the anomaly of fermion path integrals \cite{Witten2016RMP, Seiberg2016Web}. The partition function of massless fermions is not invariant under large gauge transformations due to a non-trivial spectral flow. Therefore a Chern-Simons term with a half-integer coefficient is strictly required to compensate for this anomaly, thereby ensuring the gauge invariance of the total partition function.

\textit{Renormalization group analysis and scale-invariant manifold}---
Having established the constraints on the effective field theory from the global topological data, we now turn to the quantum dynamics of the critical point itself. In the absence of the non-local term ($\lambda=0$), the pure Chern-Simons-fermion theory is known to be a conformal field theory (CFT) where the integer (or half-integer) level $k$ does not run under renormalization\cite{Wu_abelian,Wu_non_abelian}. 
A central question arises when we include the emergent non-local structure $\lambda$: does this term lead to a flow towards a new fixed point? Or, more intriguingly, does it generate a continuous family of fixed points? 
On the other hand, if charges carried by fermions or coupling can be fractional (i.e. $g=1/p$, $p$ is a positive integer), do they all flow into a given fixed point as in QED3? Or do eEFTs of fractional charges allowed by the compact gauge group all belong to a single conformal manifold (either $\lambda=0$ or $\lambda \neq 0$ ) and each allowed charge can represent a distinct CFT? 

To answer this, we perform a rigorous perturbative RG analysis towards the running parameters $k$ and $\lambda$ up to the two-loop order. Here, $g$  appears as a multiplier of charge and is quantized under compact $U(1)$ gauge group. It does not run under the RG flow. We compute the beta functions $\beta_k$ and $\beta_\lambda$ up to the two-loop order. The calculation involves evaluating vacuum polarization diagrams with internal fermion loops and non-local gauge propagators. Due to the complexity of the Feynman integrals, we employ the technique of integration by Parts (IBP). The detailed calculation and the Feynman diagrams are presented in  supplementary materials. Remarkably, through two loops we find that the beta functions vanish identically:
\begin{equation}
\beta_k(k, \lambda) = \mu \frac{dk}{d\mu} = 0, \quad \beta_\lambda(k, \lambda) = \mu \frac{d\lambda}{d\mu} = 0.
\label{eq:beta_functions}
\end{equation}
This result holds up to the two-loop order $O(k^{-2})$, and we conjecture it persists to higher orders. The vanishing of $\beta_k$ indicates that the level does not run, a feature consistent with the topological nature of the Chern-Simons term. More surprisingly, the vanishing of $\beta_\lambda$ supports the exact marginality of the nonlocal coupling and the proposed scale-invariant manifold.

For completeness, we also compute the
fermion anomalous dimension in the covariant gauge used in the Supplemental Material via a large-$k$ expansion, 
\begin{equation}
    \gamma_\psi =\frac{g^2\lambda}{3(k^2+4\pi^2\lambda^2)},
\end{equation}
which reflects interaction-induced deviations from free-fermion scaling. The critical point is not an isolated fixed point as in the standard Wilsonian paradigm but rather belongs to a continuous scale-invariant manifold parameterized by the coupling $\lambda$ as well as the discrete level $k$, charge $g$. This behavior stands in sharp contrast to standard QED$_3$ of a Maxwell form, where its infrared fixed point and dynamics depends strongly on the number of fermion flavors $N_f$.
%for small $N_f$, the theory flows to strong coupling and dynamically breaks chiral symmetry, generating a mass gap \cite{critical_qed3}. A gapless conformal theory as an isolated fixed point exists only for large $N_f$.
%In contrast to QED$_3$, 
The Chern-Simons term\footnote{The special case with $k=0$ can indeed be directly related to QED3.} here endows a topological mass to the gauge field, which screens long-range interactions and suppresses chiral symmetry breaking. This mechanism ensures that the theory remains gapless for any flavor number $N_f$, establishing a rare example of a non-compact scale-invariant manifold in $(2+1)$ dimensions realized without supersymmetry.

\textit{Wilson-loop observables}---
Despite the scale-invariant nature of the manifold, the presence of the non-local structure $\Pi_{\mu\nu}$ does have dynamical consequences.  To characterize the topological nature of the critical point, we examine the behavior of Wilson loop operators. The presence of the emergent non-local structure $\Pi_{\mu\nu}$ leads to a rich interplay between topology and dynamics.

We first consider the expectation value (VEV) of a smooth framed Wilson loop $\mathcal{W}_C=\exp(ig\oint a\cdot dx)$. Using the quadratic effective action by integrating out the fermion in Eq.(S24) and the photon propagator derived in Eq.(S1), we evaluate the path integral over $a_\mu$ explicitly. We find that the VEV receives contributions from both the non-local term and the Chern-Simons term.

The logarithm of the expectation value takes the following analytic form:
\begin{eqnarray}
   \ln \langle \mathcal{W}_C \rangle &=& - \Theta^R_{VEV}
  % \frac{g^2\lambda_{\text{eff}} }{k_{\text{eff}} ^2+4\pi^2\lambda_{\text{eff}}^2}
   \iint_C \frac{d\mathbf{x} \cdot d\mathbf{y}}{|\mathbf{x} - \mathbf{y}|^2}  
   \nonumber\\ 
   && 
  + i \, \Theta^I_{\text{VEV}}(\lambda) \cdot \text{Lk}+O(\frac{1}{k_{\text{eff}}^4}).
\label{eq:wilson_vev_formula} 
\end{eqnarray}
Here perimeter subtraction is understood for a smooth loop, and $\text{Lk}$ is the self-linking number in the chosen framing. $O(k_{\text{eff}}^{-4})$ term incorporates corrections to $\Theta^{R,I}_{VEV}$ arising from higher-order terms in $a_\mu$ induced by fermion loops as explained in supplementary materials. The exact values of the effective coupling constants can be found in Eq.(S25).  The super indices $R, I$ refer to the real and imaginary parts of the VEV of the Wilson loop operator, respectively.

The first term describes a geometry-dependent amplitude suppression.
More importantly, the second term shows that the statistical phase measured by the VEV can be further renormalized by the non-local coupling.  In the large-$k$ expansion as shown in the supplementary materials, we find explicitly that
\begin{equation}
\Theta^R_{VEV}=\frac{g^2\lambda_{\text{eff}} }{k_{\text{eff}} ^2+4\pi^2\lambda_{\text{eff}}^2},\  \  
\Theta^I_{\text{VEV}}(\lambda) = \frac{\pi  g^2 k_{\text{eff}} }{k_{\text{eff}} ^2 + 4\pi^2 \lambda_{\text{eff}} ^2}. \label{theta_vev}
\end{equation}
Thus, while the statistical phase remains a universal observable (in the sense of scale invariance), its value varies continuously along the conformal manifold. It encodes the dynamical information of the specific fixed point.

We next consider the exchange algebra of Wilson loops on a topologically non-trivial manifold, such as a torus. Let $\mathcal{W}_x$ and $\mathcal{W}_y$ be the Wilson loop operators winding around the two non-contractible cycles of the torus. As shown in the supplementary materials, the non-local term contributes to the spatial zero-mode Green function and therefore modifies the exchange angle:
\begin{equation}
\mathcal{W}_x \mathcal{W}_y
=
\exp\!\left[
i\Theta_{\mathrm{cft}}(\lambda_{\mathrm{eff}})
\right]
\mathcal{W}_y\mathcal{W}_x .
\label{eq:commutator}
\end{equation}
Within the quadratic effective theory, this angle is
\begin{equation}
\Theta_{\mathrm{cft}}(\lambda_{\mathrm{eff}})
=
2\Theta_{\mathrm{VEV}}^I.
\label{eq:theta_definition}
\end{equation}
The factor of two distinguishes the mutual exchange of two loops from the framed self-linking contribution to a single-loop VEV. Combining this with the shadowing constraint $k_{\mathrm{eff}}=k= \frac{1}{2}(c_1 + c_2)$ and defining the braiding angles of the adjacent phases $\Theta_{1,2} = 2\pi g^2/c_{1,2}$, we obtain:
\begin{equation}
\left[
\Theta_{\mathrm{cft}}(\lambda_{\mathrm{eff}})
\right]^{-1}
=
\frac{1}{2}
\left[
\Theta_1^{-1}+\Theta_2^{-1}
\right]
+
\frac{2\pi\lambda_{\mathrm{eff}}^2}
{g^2k_{\mathrm{eff}}}.
\label{eq:theta_relation}
\end{equation}
Thus, although the adjacent gapped phases fix the Chern--Simons level,
a finite $\lambda_{\mathrm{eff}}$ modifies this seemingly topological
angle, causing it to encode dynamical information about the particular
point on the scale-invariant manifold.

Eq.(\ref{eq:shadowing_result}),(\ref{eq:beta_functions}),(\ref{eq:wilson_vev_formula}),(\ref{eq:commutator}),(\ref{eq:theta_relation}) summarize our main results for the shadowing constraints, scale-invariant dynamics, and Wilson-loop
observables at these tQCPs.
The significance of Eq.~(\ref{eq:theta_relation}) becomes clear when compared with standard QCP occurring in symmetry-protected topological states (SPTs) \cite{Grover_tQCP, LuLee2014, Tsui2015}. In typical quantum phase transitions in SPTs, the infrared degrees of freedom as well as their dynamics are determined solely by the changes of topological invariants $\delta N_w$ and protecting symmetries $G_p$. For instance in a few recent studies of 3D and 2D SPTs\cite{Zhou2022,Zhou2024,Yang2025,Meyniel2025}, it was illustrated explicitly that the number of gapless flavors $N_f$ follows a relation of the form:
\begin{equation}
N_f = N_f^0(G_p) \frac{\delta N_w}{\delta N_w^0(G_p)},
\label{eq:spt_relation}
\end{equation}
where $N_f^0(G_p) $ is the degrees of freedom in a fundamental field representation of $G_p$. In addition, $\delta N_w^0(G_p)$ is the
change of topological invariants in the fundamental representation,
again fixed by the global symmetry $G_p$. Crucially, for a given protecting symmetry, all tQCPs belong to the same universality class once $\delta N_w$ is further fixed. They are independent of specific values of $N_w$ at which topological quantum phase transitions take place. 

%In our case, the physics is richer in two aspects. First, due to topological shadowing, the universality class depends explicitly on the absolute values of the chiral charges $c_1$ and $c_2$. Second, within each universality class defined by $k$, there exists a scale-invariant manifold parameterized by $\lambda$, allowing for continuous variation of dynamical properties (like $\Theta_{\text{VEV}}$) while the topological algebra and commutator remain fixed.

\textit{Holographic perspective}---Before concluding, we discuss a complementary, holographic perspective on the origin of the non-local $\lambda$-term, which, interestingly, also sheds extra light on the exact marginality of the non-local term. 
%The non-local term of our $(2+1)$D theory can naturally emerge on the boundary of a $(3+1)$D bulk system. We can interpret this extra term as arising from a graphene-like sheet in two spatial dimensions coupled to a dynamical photon in three spatial dimensions~\cite{Son:2007ja}.
This holographic perspective on the exact marginality of the non-local term was first extensively discussed by Di Pietro, Gaiotto, et al.~\cite{DiPietro:2019hqe}. Within this framework, the effective self-energy term shown in Eq.~\eqref{eq:non_local_term} can be interpreted as the boundary description of a $(3+1)$D free Maxwell theory with a topological $\theta$-term.
\begin{equation}\label{eq:4d}
\mathcal{L}_{\text{4d}} = \frac{1}{4g^2}F_{\mu\nu} F^{\mu\nu} + \frac{i\theta}{32\pi^2}\epsilon^{\mu\nu\rho\sigma}F_{\mu\nu}F_{\rho\sigma}.
\end{equation}
%Upon integrating out the photon degrees of freedom in the extra dimension, we obtain the effective self-energy term shown in Eq.~\eqref{eq:non_local_term}.
%In a related context, Eq.~\eqref{eq:non_local_term} also naturally emerges when the fermion theory is realized on the boundary of a $(3+1)$-dimensional Maxwell theory~\cite{DiPietro:2019hqe} with the Lagrangian given in Eq.~\eqref{eq:4d}. 
%\begin{equation}
%\mathcal{L}_{\text{4d}} = \frac{1}{4g^2}F_{\mu\nu} F^{\mu\nu} + \frac{i\theta}{32\pi^2}\epsilon^{\mu\nu\rho\sigma}F_{\mu\nu}F_{\rho\sigma}.
%\end{equation}
After integrating out the contribution from the extra dimension, the propagator of the gauge boson takes the form\footnote{The gauge-invariant correlator $\langle F_{\mu\nu}(p)F_{\rho\sigma}(-p) \rangle$ is computed first in~\cite{DiPietro:2019hqe}. Extracting the gauge field propagator $\langle A_{\mu}(p)A_{\nu}(-p) \rangle$ introduces a gauge-fixing parameter $\xi$, modifying the first term to $[\delta_{\mu\nu} - (1-\xi)p_{\mu}p_{\nu}/p^2]/p$. Here we present the $\xi=1$ (Feynman) gauge to facilitate direct comparison with Eq.~\eqref{eq:propagator_app}.}
\begin{equation}
G_{\mu\nu}(p)=\langle A_\mu(p) A_\nu(-p)\rangle 
=\frac{g^2}{1 + \gamma^2}(\frac{\delta_{\mu\nu}}{p} + \gamma\frac{\epsilon^{\mu\nu\lambda}p_\lambda}{p^2}), \label{eq:propagator_4d}
\end{equation}
where $\gamma = \frac{g^2\theta}{4\pi^2}$. Comparing it with Eq.(S1), we see that the two propagators are identical when we tune the bulk parameter $g^2$ to be equal to $1/\lambda$. Since the bulk is a free theory and the Maxwell term $F_{\mu\nu}F^{\mu\nu}$ is an exactly marginal parameter, it is a first hint that the non-local term is also exactly marginal with vanishing beta function.
As argued in~\cite{DiPietro:2019hqe}, the boundary non-local term is expected to be exactly marginal because it inherits the scale invariance of the free $(3+1)$D bulk. Our two-loop analysis provides a rigorous dynamical verification of this conjecture in the presence of $N_f$ flavors of Dirac fermions. The identically vanishing $\beta_k$ and $\beta_\lambda$ confirm that this exact marginality is a robust feature of the $(2+1)D$ eEFT, likely persisting to all loop orders.
 %This correspondence offers a profound insight: since the $(3+1)$D bulk is a free Maxwell theory and the topological $\theta$-parameter is also exactly marginal coupling, it provides a compelling hint that the boundary non-local term must also be exactly marginal with a vanishing beta function. Indeed, our results are perfectly consistent with this holographic intuition. Although our renormalization group analysis is explicitly carried out up to the two-loop level, this deep structural bulk-boundary connection leads us to conjecture that the vanishing of the beta functions holds to arbitrary loop orders. 
 Collectively, these scenarios demonstrate that this non-local interaction can also be highly relevant to boundaries of a physical system implying a possible holographic representation of our eEFT and holographic nature of the tQCP of our interest\footnote{In principle, an eEFT can always have a nonzero $\lambda$ when an additional background sector of gapless degrees of freedom happens to be present near tQCPs as a bystander but does not drive the specific topological transition which we are interested in.}. We discuss the holographic perspective in more detail from the point of view of $(-1)$-form symmetry in the supplementary material.

\textit{Conclusion and discussions}---
In this work, we have proposed a theoretical framework for topologically shadowed tQCPs, establishing a rigorous correspondence between topological data of the gapped phase and critical dynamics. Our study yields two main insights. First, we introduced the principle of topological shadowing, where the defining parameters of the IR critical theory can be uniquely constrained by the chiral charges of the adjacent gapped phases and their anyon topological data (i.e., UV data). 
%This anomaly-matching constraint manifests in a universal relation of the topological angle at a critical point, $\Theta_{\text{cft}}^{-1} = (\Theta_1^{-1} + \Theta_2^{-1})/2$. 
This anomaly-matching constraint fixes the discrete critical-theory
parameters through $k=(c_1+c_2)/2$ and $N_f=c_2-c_1$. This distinguishes these transitions from standard phase transitions such as those in SPTs. Second, we found that quantum critical dynamics here are characterized by a non-compact scale-invariant manifold parameterized by the non-local term coupling $\lambda$.  This provides a rare example of a continuous manifold of fixed points in $(2+1)$D realized without supersymmetry. Dynamical observables, including the fermion anomalous dimension and the Wilson-loop exchange angle $\Theta_{\mathrm{cft}}$, vary continuously along this manifold, while the discrete shadowing constraints on $k$ and $N_f$ remain fixed.

%While dynamical observables vary continuously along this manifold, we showed that the global topological algebra remains a robust invariant. 
%It was also recently suggested that conformal manifolds can be a powerful tool to understand structures of overwhelmingly or even exponentially large numbers of isolated fixed points that can naturally appear when the interaction parameter space is high-dimensional at tQCP\cite{Saran2025,Saran2026}. 

Several future directions are in order. We anticipate that our tQCP framework can provide a potential field-theoretical description for tunable quantum phase transitions between Fractional Chern Insulator (FCI) states in twisted graphene platforms\cite{2021Natur.600..439X,2024Natur.626..759L,PhysRevLett.122.106405,PhysRevB.108.085117}. While flat-band dynamics drive these transitions, contributions from other bands touching the Fermi level with higher velocities may not be neglected, as discussed below Eq.~\eqref{eq:non_local_term}. Integrating out these fermion fields
may also lead to a similar non-local $\lambda$-term to the eEFT for the flat band. This hints the relevance of our theory in an experimental setup. Theoretically, if we assume emergent \emph{conformal} invariance rather than mere scale invariance, we can compare our calculated scaling dimensions with those obtained via the fuzzy sphere \cite{Zhu:2022gjc,Zhou:2025rmv} or conformal bootstrap \cite{Chester:2016wrc}, a very interesting avenue that warrants further investigation.

\textit{Acknowledgments}--- We thank Chunxiao Liu, Chong Wang, and Zheng Zhou for discussions regarding the origin of the nonlocal $\lambda$ term. This work is in part supported by funding from Hong Kong’s Research Grants Council (GRF No.14307624, RFS2324-4S02 and AoE/P-404/18). WY is supported by the Natural Sciences and Engineering Research Council of Canada (NSERC) and the European Commission under the Grant Foundations of
Quantum Computational Advantage. FZ is supported by an NSERC (Canada) Discovery Grant under contract No RGPIN-2020-07070.

\section{Two-Loop Beta Function Calculation}
\label{sm:beta_function}
\renewcommand{\theequation}{S\arabic{equation}}

\begin{figure}[htbp]
\centering
\includegraphics[scale=0.05]{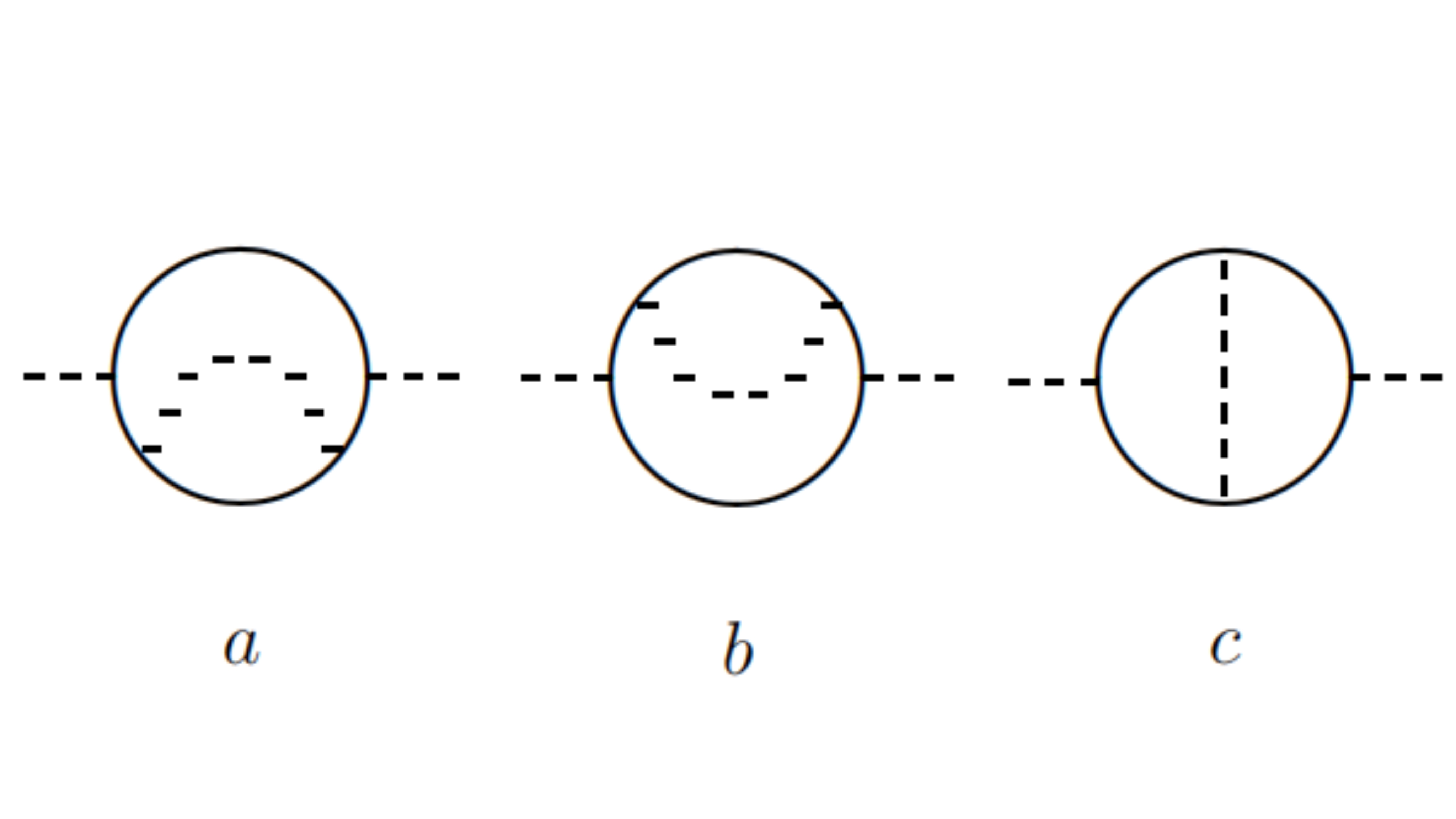}
\caption{Two-loop diagrams of photon self-energy. Solid lines denote fermion propagators, while dashed lines denote gauge field propagators.}
\label{abc}
\end{figure}

In this appendix, we present the calculation of the beta functions for the coupling constants $k$ and $\lambda$. We start with the effective action Eq.(\ref{eq:lagrangian}). 
 We take the Landau gauge, by adding a gauge fixing term $\frac{(\partial_\mu a^\mu)^2 }{2\xi\sqrt{-\partial^2}}$ to the Lagrangian and then take  $\xi=\frac{4\pi^2\lambda}{k^2+4\pi^2 \lambda^2}$ in the resulting propagator for the gauge field, which yields
\begin{eqnarray}
G_{\mu\nu}(p)&=&\langle a_\mu(p) a_\nu(-p)\rangle \nonumber \\
&=&\frac{4\pi^2}{k^2+4\pi^2 \lambda^2}(\frac{\lambda\delta_{\mu\nu}}{p}-\frac{k}{2\pi}\frac{\epsilon^{\mu\nu\lambda}p_\lambda}{p^2}). \label{eq:propagator_app}
\end{eqnarray}
The gauge invariance implies that the polarization tensor of photon can be written as 
\begin{eqnarray}
\Pi_{\mu\nu}=\Pi_e(p)p(\delta_{\mu\nu}-\frac{p_\mu p_\nu}{p^2})+\Pi_o(p)\epsilon_{\mu\nu\rho}p^\rho. \label{polarization}
\end{eqnarray}
Here subscripts $o$ and $e$ denote the P-odd and P-even components respectively. The one-loop  polarization tensor due to the massless fermion bubble is finite by using dimensional regularization ($D=3-\epsilon$) and purely symmetric in the indices, given by:
\begin{eqnarray}
    \Pi_e^{(1)}=-\frac{g^2N_f}{16},\ \ \Pi_o^{(1)}=0.
\end{eqnarray}
We evaluate the two-loop vacuum polarization diagrams (photon self-energy) shown in Fig.~\ref{abc}.  For the P-odd  components, the contributions from the diagrams $(a+b)$ and $(c)$ are given by:
\begin{widetext}
\begin{align}
\Pi_{o(a+b)}^{(2)} &=  \frac{16\pi k g^4N_f}{k^2 + 4\pi^2\lambda^2} \int \frac{d^{D} p}{(2\pi)^{D}} \frac{d^D q}{(2\pi)^D} \frac{[l \cdot (p+l)] [q \cdot (p-q)]}{p^2 (p+l)^2 (p-q)^2 q^2 l^2}, \label{eq:pi_odd_ab_raw} \\
\Pi_{o(c)}^{(2)} &= \frac{4\pi k g^4N_f}{k^2 + 4\pi^2\lambda^2}  \int \frac{d^{D} p}{(2\pi)^{D}} \frac{d^D q}{(2\pi)^D} \frac{\mathcal{F}(q,p,l)}{p^2 (p+l)^2 (p-q)^2 q^2 (q+l)^2l^2},
\label{eq:pi_odd_c_raw}\\
\mathcal{F}(q,p,l) &= l^2 \Big[ 2 p^2 q^2 - 4 (p \cdot q)^2 + (p^2 + q^2)(p \cdot q) + (q^2 - p \cdot q)(p \cdot l) + (p^2 -     p \cdot q)(q \cdot l) \Big] \nonumber \\
&\quad - q^2 (p \cdot l)^2 - p^2 (q \cdot l)^2- \big( p^2 + q^2 - 4 p \cdot q \big)(p \cdot l)(q \cdot l).
\end{align}

For the P-even  components, the contributions from the diagrams $(a+b)$ and $(c)$ are given by:
\begin{align}
\Pi_{e(a+b)}^{(2)} &=\frac{8\pi^2 \lambda g^4N_f}{k^2 + 4\pi^2\lambda^2} \int \frac{d^{D} p}{(2\pi)^{D}} \frac{d^D q}{(2\pi)^D}  \frac{1}{ p^4 (p+l)^2(p-q)^2 ql} \Big[ 2(l \cdot p)(p \cdot q) - p^2 \big( l \cdot (p+q) + p^2 - p \cdot q \big) \Big], \\
\Pi_{e(c)}^{(2)} &= \frac{4\pi^2 \lambda g^4N_f}{k^2 + 4\pi^2\lambda^2} \int \frac{d^{D} p}{(2\pi)^{D}} \frac{d^D q}{(2\pi)^D} \frac{1}{ p^2 q^2 (p+l)^2(q+l)^2 |p-q|l} \Big[ 4(p \cdot q)(l \cdot p + l \cdot q - p \cdot q) + 4(l \cdot p)(l \cdot q) \nonumber \\
& \quad \quad - (p \cdot q)l^2 - (l \cdot p)q^2 - (l \cdot q)p^2 - p^2 q^2 \Big].
\end{align}
\end{widetext}

The momentum integration can be evaluated by using  the IBP technique \cite{IBP_review}. This method allows us to reduce the complicated two-loop scalar integrals into products of one-loop master integrals.

We first define the standard one-loop master integral $I(a,b)$ in $D$-dimensional Euclidean space:
\begin{align}
I(a,b) &= \int \frac{d^D p}{(2\pi)^D} \frac{1}{p^{2a}(p-k)^{2b}} \nonumber \\
&= C(a,b) k^{-2(a+b-D/2)},
\label{eq:one_loop_master}
\end{align}
where the coefficient function $C(a,b)$ is given by:
\begin{equation}
\begin{aligned}
C(a,b) &= \frac{1}{(4\pi)^{D/2}} \frac{\alpha(a)\alpha(b)}{\alpha(a+b-D/2)}, \\
\alpha(x) &= \frac{\Gamma(D/2-x)}{\Gamma(x)}.
\end{aligned}
\end{equation}

The generic scalar two-loop integral $S(a,b)$ appearing in our self-energy calculation takes the form:
\begin{align}
S(a,b) &= \int \frac{d^D p}{(2\pi)^D} \int \frac{d^D q}{(2\pi)^D} \nonumber \\
&\quad\times \frac{1}{p^{2a}(p+k)^{2b}(p-q)^2}\frac{1}{(q+k)^2 q^2} \nonumber \\
&= J(a,b) k^{-2(a+b+3-D)}.
\label{eq:two_loop_scalar}
\end{align}
To compute this, we utilize the IBP identity based on the vanishing of the integral of a total derivative in dimensional regularization:
\begin{equation}
\int \frac{d^D p}{(2\pi)^D} \frac{\partial}{\partial p_\mu} \frac{(p-q)_\mu}{p^{2a}(p+k)^{2b}(p-q)^2(q+k)^2 q^2} = 0.
\label{eq:ibp_identity}
\end{equation}
Applying the derivative and performing algebraic manipulations, we obtain a recurrence relation that expresses the two-loop integral $S(a,b)$ entirely in terms of the one-loop function $I(a,b)$:
\begin{align}
J(a,b) = & \frac{C(1,1)}{D-a-b-2} \nonumber \\
\times\Big[ & a C(a+1, b) - a C(1+a, b+2-\frac{D}{2}) \nonumber \\
+ & b C(1+b, a) - b C(1+b, a+2-\frac{D}{2}) \Big].
\label{eq:ibp_reduction}
\end{align}
Using Eq.~(\ref{eq:ibp_reduction}) in conjunction with Eq.~(\ref{eq:one_loop_master}), we can analytically evaluate all contributions to the polarization tensor.

The contributions to the pole part of polarization tensor components from the relevant diagrams  are found to be\footnote{Our result deviates slightly from Ref.~\cite{Wu_abelian,Wu_non_abelian}. The origin of this difference lies in their treatment of the two-loop integrals: one momentum is integrated  out first with fixed dimension $3$, whereas we employ the IBP method with both  momentum integrals evaluated in $D$ dimensions. Consequently, the final results  differ by a factor of $1/2$.}:
\begin{align}
\Pi_{o(a+b)}^{(2)} &= -\frac{g^4 N_f}{24\pi^2\epsilon} \frac{2\pi k}{k^2 + 4\pi^2\lambda^2}, \nonumber \\
\Pi_{o(c)}^{(2)} &= \frac{g^4 N_f}{24\pi^2\epsilon} \frac{2\pi k}{k^2 + 4\pi^2\lambda^2}, \nonumber \\
\Pi_{e(a+b)}^{(2)} &= \frac{g^4 N_f}{48\pi^2\epsilon} \frac{4\pi^2 \lambda}{k^2 + 4\pi^2\lambda^2}, \nonumber \\
\Pi_{e(c)}^{(2)} &= -\frac{g^4 N_f}{48\pi^2\epsilon} \frac{4\pi^2 \lambda}{k^2 + 4\pi^2\lambda^2}.
\end{align}

Crucially, the pole parts cancel separately in the parity-even and parity-odd sectors:
\begin{equation}
\Pi_{\text{total}}^{(2)} = \Pi_{(a+b)}^{(2)} + \Pi_{(c)}^{(2)} = 0.
\end{equation}
To systematically analyze the renormalization group flow, it is convenient to absorb the Chern-Simons level $k$ into the gauge field by the rescaling $a_\mu \to a_\mu/\sqrt{k}$. In this rescaled frame, the renormalized action including the counterterms is specified as:
\begin{eqnarray}
 S_{\text{ren}} = \int d^3x \bigg[ &&\sum_{i=1}^{N_f} Z_\psi \bar{\psi}_i \gamma^\mu \left(\partial_\mu - i \frac{Z_2}{Z_\psi}\eta a_\mu \right)\psi_i \nonumber\\
    &&+ \frac{i Z_a}{4\pi} \epsilon^{\mu\nu\rho} a_\mu \partial_\nu a_\rho + Z_3\frac{ \lambda}{2k} a_\mu \hat{\Pi}_{\mu\nu} a_\nu \bigg],\nonumber\\
\label{renormalized_action}
\end{eqnarray}
where $\eta=g/\sqrt{k}$. The Ward identity\footnote{Dimensional regularization ensures gauge invariance at the two-loop level, 
and consequently the Ward identity holds.}  requires $Z_\psi = Z_2$, which implies that the vertex renormalization cancels the fermion wave-function renormalization. By defining $\eta=Z_\eta \eta_{\text{bare}},\, \lambda=Z_\lambda \lambda_{\text{bare}}$ and $k=Z_k k_{\text{bare}}$, we can construct them using the renormalization constant listed in Eq.(\ref{renormalized_action}):
\begin{eqnarray}
    Z_2=Z_\eta Z_\psi (Z_a)^{1/2},\ \ Z_3=Z_a Z_\lambda/Z_k.
\end{eqnarray}
The Ward identity dictates the relation between the coupling renormalization constant and the gauge field renormalization constant $Z_a$:
\begin{equation}
    Z_\eta = Z_a^{-1/2},\ \ Z_\lambda=Z_3Z_k/Z_a.
\end{equation}
Since our calculation shows that $Z_3,\,Z_a$ are finite and $g$ is the multiplier of charge, there is no anomalous dimension associated with the coupling. Consequently, the beta functions vanish identically:
\begin{equation}
\beta_k(k,\lambda) = 0, \quad \beta_\lambda(k,\lambda) = 0.
\end{equation}
It is important to emphasize that while the effective coupling strength $\eta$ receives a finite renormalization due to the finite part of $Z_a$ (effectively shifting $\eta$), this finite correction depends on the regularization scheme and should be regarded as an artifact rather than a running of the quantized parameters $g$ and $k$.

\begin{figure}[htbp]
\centering
\includegraphics[scale=0.6]{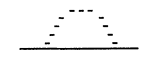}
\caption{One-loop diagrams of fermion self-energy.}
\label{fse}
\end{figure}

To explicitly demonstrate that the critical point is a genuinely interacting conformal field theory rather than a free theory, we calculate the fermion anomalous dimension $\gamma_\psi$. Unlike pure Chern-Simons theory where the one-loop fermion self-energy is finite, the presence of the non-local term introduces a logarithmic divergence at the one-loop level. At one-loop order it is given by the diagram in Fig.~\ref{fse},  we obtain:
\begin{equation}
\Sigma^{(1)}(p) =- i \slashed{p} \frac{2g^2\lambda}{3(k^2+4\pi^2\lambda^2)}\frac{1}{\epsilon}+\text{finite term $\propto p$} .
\label{eq:sigma_1loop}
\end{equation}
Notice that at this order the renormalized mass remains zero, $\Delta M = \Sigma(p=0) = 0$, which is consistent with the massless critical theory. The inverse of the full fermion propagator is given by $S^{-1}(p) = i\slashed{p} - \Sigma(p)$. The fermion wave function renormalization constant $Z_\psi^{-1}$ is extracted from the divergent part of the kinetic term:
\begin{equation}
Z_\psi^{-1} = \left[ \frac{\partial S^{-1}(p)}{\partial (i\slashed{p})} \right]_{p=0} = 1 + \frac{2g^2\lambda}{3(k^2+4\pi^2\lambda^2)}\frac{1}{\epsilon}.
\label{eq:Z_psi_inv}
\end{equation}
 One can thereby obtain the anomalous dimension $\gamma_\psi$ of the massless fermion field up to the first order:
\begin{equation}
\gamma_\psi = -\frac{1}{2} \frac{\partial Z_\psi}{\partial (1/\epsilon)} = \frac{g^2\lambda}{3(k^2+4\pi^2\lambda^2)}.
\label{eq:gamma_psi}
\end{equation}

Finally, to systematically derive the effective dynamics for the gauge field observables (such as the Wilson loop VEV discussed in the main text), we consider the effective action $S_{\text{eff}}[a]$ obtained by integrating out the fermion fields $\psi$ while treating the gauge field $a_\mu$ as a background field. The partition function can be written as $Z = \int \mathcal{D}a \, e^{-S_{\text{eff}}[a]}$, where:
\begin{equation}
S_{\text{eff}}[a] = S_{\text{CS}}[a]  +S_\lambda[a]
-N_f \ln \det \gamma_\mu (\partial_\mu -ig a_\mu).
\end{equation}
Here $S_\lambda[a]$ is the bare nonlocal action, and the determinant is for one flavor. Expanding the fermion determinant in powers of $a_\mu$, we obtain the quadratic term along with an infinite series of higher-order gauge-invariant interaction terms:
\begin{eqnarray}
     S_{\text{eff}}[a] &= &\frac{\lambda_{\text{eff}}}{2} \int \frac{d^3p}{(2\pi)^3} a_\mu(-p) \hat{\Pi}_{\mu\nu}(p) a_\nu(p)\nonumber\\
    &&  + \frac{ik_{\text{eff}}}{4\pi} \int d^3x \epsilon_{\mu\nu\rho} a_\mu \partial_\nu a_\rho  +  \sum_{n=3}^{\infty} S^{(n)}_{\text{int}}[a],\nonumber\\
    \label{gauge_field_eff}
\end{eqnarray}
where the non-local coefficient $\lambda_{\text{eff}}$ includes the finite one-loop correction:
\begin{equation}
    \lambda_{\text{eff}} = \lambda + \frac{g^2 N_f }{16}, \quad k_{\text{eff}} = k.
    \label{effective_coupling_const}
\end{equation}
Note that $\lambda_{\text{eff}}=g^2 N_f /16$ for the critical dynamics when $\lambda=0$. At $m=0$, the one-loop polarization shifts only the parity-even coefficient in our convention. The higher-order terms $S^{(n)}_{\text{int}}$ (e.g., boson-boson scattering vertices induced by fermion box diagrams) are of order $\mathcal{O}(a^n)$. When calculating correlation functions of the gauge field, such as the Wilson loop expectation value, these terms are treated as interaction vertices.
It is crucial to note that these higher-order corrections do not alter the form of Eq.(\ref{eq:wilson_vev_formula}). Due to gauge invariance, the vacuum polarization tensor describing the correlation of the current sources ($J^{\mu}(x)=\oint dy^{\mu}\delta^{3}(x-y)$) must take the form of Eq.(\ref{polarization}). Consequently, the interaction effects induced by fermion loops only result in a finite, scale-independent renormalization of the coefficients $\Theta_{\text{VEV}}^R$ and $\Theta_{\text{VEV}}^I$, without introducing new non-local structures. The explicit analytic forms given in Eq.~(\ref{eq:wilson_vev_formula}) therefore remain valid, with the effective parameters receiving finite corrections at higher orders in the large $1/k$ expansion.

\section{Exchange algebra of torus Wilson loops}
\label{sm:commutator_proof}

In this appendix, we calculate how the non-local structure modifies
the exchange algebra of Wilson loops on a torus $T^2$. We work in the Lorenz gauge $\partial_\mu a_\mu=0$.

The gauge field $a_\mu(\mathbf{x}, t)$ can be expanded using a complete basis of Fourier modes:
\begin{equation}
a_\mu(\mathbf{x}, t) = \frac{1}{L} \sum_{\mathbf{n}} q_{\mu, \mathbf{n}}(t) \, e^{i \mathbf{k_n} \cdot \mathbf{x}}, \quad \mathbf{k_n} = \frac{2\pi}{L}(n_x, n_y).
\end{equation}
Here, $q_{\mu, \mathbf{n}}(t)$ are time-dependent coefficients. The Lorenz gauge condition implies the constraint 
\begin{equation*}
    \dot{q}_{0, \mathbf{n}}+\frac{2\pi i}{L}n_x q_{x, \mathbf{n}}+\frac{2\pi i}{L}n_y q_{y, \mathbf{n}}=0.
\end{equation*}
For the spatial zero mode $\mathbf n=\mathbf 0$, this condition gives $\dot q_{0,\mathbf 0}=0$, while the spatial holonomies $q_{x,\mathbf 0}$ and $q_{y,\mathbf 0}$ remain dynamical.

Consider the Wilson loop operators winding around the two non-contractible cycles of the torus, $\mathcal{W}_x$ along the $x$-direction and $\mathcal{W}_y$ along the $y$-direction. Substituting the mode expansion into the line integrals, we have:
\begin{align}
\mathcal{W}_x &= \exp( i g \int_0^L dx\, a_x ) = \exp [ i g \sum_{n_y} q_{x, (0, n_y)}(t)], \nonumber\\
\mathcal{W}_y &= \exp ( i g \int_0^L dy\, a_y  ) = \exp [ i g \sum_{n_x} q_{y, (n_x, 0)}(t)].
\end{align}
The commutator $[\mathcal{W}_x, \mathcal{W}_y]$ depends on the commutation relations of these mode coefficients. Within the quadratic approximation to the effective action, modes with different spatial momenta decouple by momentum conservation. The cross correlator relevant to $\mathcal W_x\mathcal W_y$ therefore reduces to the spatial zero-mode sector.

We denote the spatial zero modes by
$q_i\equiv q_{i,\mathbf 0}$, where $i=x,y$, and use
$q_i(\tau)=\int d\omega\,e^{i\omega\tau}q_i(\omega)/(2\pi)$.
The quadratic Euclidean zero-mode action is
\begin{equation}
S_{\mathrm{quad}}^{(0)}
=
\frac{1}{2}
\int_{-\infty}^{\infty}\frac{d\omega}{2\pi}\,
q_i(-\omega)
\left[
\lambda_{\mathrm{eff}}|\omega|\delta_{ij}
+
\frac{k_{\mathrm{eff}}}{2\pi}\omega\epsilon_{ij}
\right]
q_j(\omega).
\end{equation}
Although the non-local kernel is symmetric in the spatial indices,
it contributes to the inverse quadratic kernel.  Its antisymmetric
component is
\begin{equation}
G_{xy}^{(0)}(\omega)
=
-\frac{2\pi k_{\mathrm{eff}}}
{k_{\mathrm{eff}}^2+4\pi^2\lambda_{\mathrm{eff}}^2}
\lim_{\varepsilon\to0^+}
\frac{\omega}{\omega^2+\varepsilon^2}. \label{Gomega}
\end{equation}
Fourier transforming Eq.~(\ref{Gomega}), we obtain
\begin{equation*}
G_{xy}^{(0)}(\tau)
=
-\frac{\pi i k_{\mathrm{eff}}}
{k_{\mathrm{eff}}^2+4\pi^2\lambda_{\mathrm{eff}}^2}
\operatorname{sgn}(\tau).
\end{equation*}
For the Euclidean time-ordered Green function
$G_{xy}^{(0)}(\tau)
=\langle T_\tau q_x(\tau)q_y(0)\rangle$, its discontinuity gives
\begin{equation}
[q_{x,\mathbf 0},q_{y,\mathbf 0}]
=
G_{xy}^{(0)}(0^+)-G_{xy}^{(0)}(0^-)
=
-\frac{2\pi i k_{\mathrm{eff}}}
{k_{\mathrm{eff}}^2+4\pi^2\lambda_{\mathrm{eff}}^2}.
\end{equation}
Since this commutator is a $c$-number within the quadratic theory,
the Baker--Campbell--Hausdorff formula gives
\begin{equation}
\mathcal W_x\mathcal W_y
=
\exp\!\left[
i\frac{2\pi g^2 k_{\mathrm{eff}}}
{k_{\mathrm{eff}}^2+4\pi^2\lambda_{\mathrm{eff}}^2}
\right]
\mathcal W_y\mathcal W_x .
\end{equation}
Thus, the exchange angle of the torus Wilson loops depends on
$\lambda_{\mathrm{eff}}$ and reduces to the pure Chern--Simons result
only in the formal limit $\lambda_{\mathrm{eff}}=0$.

\section{A perspective from \texorpdfstring{$(-1)$}{(-1)}-Form Symmetry}\label{sm:-1_form}

In this appendix, we review the holographic perspective from the angle of generalized global symmetries \cite{Gaiotto_2015}. 

A $p$-form global symmetry in $D$ dimensions is generated by a $(D-p-1)$-form conserved current. In $(3+1)$-dimensional spacetime, the free Maxwell theory in Eq.~\eqref{eq:4d} possesses two distinct exactly marginal deformations, one corresponding to the Maxwell term and another corresponding to the theta term. In the framework of generalized global symmetries, these marginal operators can be formally identified as the conserved currents of $(-1)$-form symmetries. The coupling constants for these operators serve as the 0-form background gauge fields. 

Specifically, the $(-1)$-form symmetry associated with the $\theta$-term is a topological $U(1)$ $(-1)$-form symmetry. The conserved current of the $(-1)$-form symmetry is the topological density,
    \begin{equation}
        J_{\text{top}}^{(4)} = \frac{1}{32\pi^2}\epsilon^{\mu\nu\rho\sigma}F_{\mu\nu}F_{\rho\sigma}.
    \end{equation}
The integral of $J_{\text{top}}^{(4)}$ over a closed 4-manifold measures the instanton number, which is topologically quantized to integer values. Hence, this symmetry arises from the topological instanton density of a \textit{compact} $U(1)$ gauge theory. Because the charge is quantized, shifting the associated background field $\theta \to \theta + 2\pi$ leaves the path integral weight $e^{i \int \mathcal{L}_{4d}}$ invariant. Thus, the symmetry group is a compact $U(1)$. The charged objects are $(-1)$-dimensional defects, physically corresponding to instantons (localized events in space and time).

On the other hand, the $(-1)$-form symmetry associated with the exactly marginal Maxwell term is a dynamical $\mathbb{R}$ $(-1)$-form symmetry. The conserved current is the Maxwell term itself,
\begin{equation}
    J_{\text{dyn}}^{(4)} = F_{\mu\nu} F^{\mu\nu}.
\end{equation}
Unlike the topological $(-1)$-form symmetry, $J_{\text{dyn}}^{(4)}$ is manifestly dependent on the background geometry. The integrated charge evaluates to the classical action, which is a continuous, positive real number. Consequently, the symmetry group is the non-compact reals, $\mathbb{R}$. Acting with this $\mathbb{R}$ $(-1)$-form symmetry scales the partition function weight and alters correlation functions. Rather than representing a pure topological invariant, shifting this background field represents moving along the \textit{conformal manifold} of the $(3+1)$D theory, shifting the system to a different conformal field theory within the parameter space.

Now consider the boundary of this $(3+1)$D theory. Going from the bulk to the boundary, the current $J_{\text{dyn}}^{(4)}$ becomes exactly the non-local $\lambda$ term in Eq.~\eqref{eq:non_local_term}. Hence, we can think of this non-local term as coming from the operator $J^{(4)}$ restricted to the boundary, analogous to the ``tilt'' operator for an ordinary 0-form symmetry \cite{Billo:2016cpy,Gimenez-Grau:2022czc}. Although this operator is non-local, the general structure of conformal field theories still suggests that this non-local term is exactly marginal, as confirmed by our two-loop calculation. It is interesting to see whether further information can be obtained from this perspective.

In passing, we mention that because free Maxwell theory possesses exact $S$-duality, these two $(-1)$-form symmetries are naturally unified. The two 0-form background fields ($\theta$ and $1/g^2$) combine into a single complexified coupling constant:
\begin{equation}
    \tau = \frac{\theta}{2\pi} + \frac{4\pi i}{g^2}
\end{equation}
This parameter $\tau$ acts as a complex background scalar field. It couples to the self-dual and anti-self-dual combinations of the field strength, effectively generating a complexified $(-1)$-form symmetry that governs the conformal manifold $\mathbb{H}/\mathrm{SL}(2, \mathbb{Z})$ of the $(3+1)$D free Maxwell theory.

\bibliography{ref}

\end{document}